# The Dublin Lens: A Cc=1.0 mm Objective Lens Intended for CryoEM at 100 keV

Germano Motta Alves[1,2], Theo Andrews[1,2], Patrick McBean[1], Torquil Wells[3], Mohamed M. El-Gomati[3], Stephan Burkhalter[4], Clemens Schulze-Briese[4], Pietro Zambon[4], Greg McMullan[*5], Richard Henderson[5], Christopher J. Russo[5], Lewys Jones[*1,2]

**Affiliations**

1. School of Physics & Advanced Microscopy Laboratory (CRANN), Trinity College Dublin, Dublin, Ireland
2. turboTEM Ltd, Dublin, Ireland
3. York Probe Sources Ltd., York YO26 6QU, United Kingdom
4. DECTRIS Ltd. Baden-Daettwil, Switzerland
5. MRC Laboratory of Molecular Biology, Francis Crick Avenue, Cambridge CB2 0QH, United Kingdom

\* Corresponding authors: lewys.jones@tcd.ie, gm2@mrc-lmb.cam.ac.uk

**Abstract**

We have designed, fabricated and tested a lens with chromatic aberration coefficient (Cc) of 1.0 mm, a 4.0 mm pole-gap and 2.0 mm bore that is wide enough to accommodate an anti-contamination system and an objective aperture. This lens extends the temporal-coherence envelope of the electron microscope beyond 2 Å, using a low-cost Schottky FEG. We hope that this lens design can be used to improve all 100 keV electron microscopes designed for single-particle electron cryomicroscopy (cryoEM).

## Introduction

The primary method for determining the structure of biological molecules has recently become single-particle cryoEM [1]. Previous work has shown that 100 keV electrons have the potential to improve structure determination by cryoEM since there is more information in the images at 100 keV per unit damage than at higher energies [2],[3]. Recent work has demonstrated that structure determination by cryoEM at 100 keV is both possible and practical [4],[5],[6],[7]. Furthermore, two hardware modifications were identified in [4] that can offer improvement in microscope performance with minimal increases in cost:

1. Modification of the objective lens to reduce the effects of chromatic aberration which increases the signal at high resolution.
2. Modification of the detector to increase the number of pixels such that the detective quantum efficiency at a particular resolution can be increased using magnification.

In this work, we address the first of these. Given the growth of cryoEM and the ever increasing cost of high-end microscopes, there is still a desperate need for affordable microscopes capable of structure determination and walk-up assessment of specimens. Interestingly, improving the objective lens performance does not increase the complexity of the microscope and so has the potential to enable higher performance without increasing cost.

Transmission electron microscope (TEM) objective lenses consist of a (typically) copper coil which produces the excitation, a 'yoke' which guides the flux to complete the magnetic circuit, and a pole-piece with a fixed bore and gap that dictate the shaping of the magnetic field in the focussing region. A simplified illustration of such a lens is shown in Figure 1.

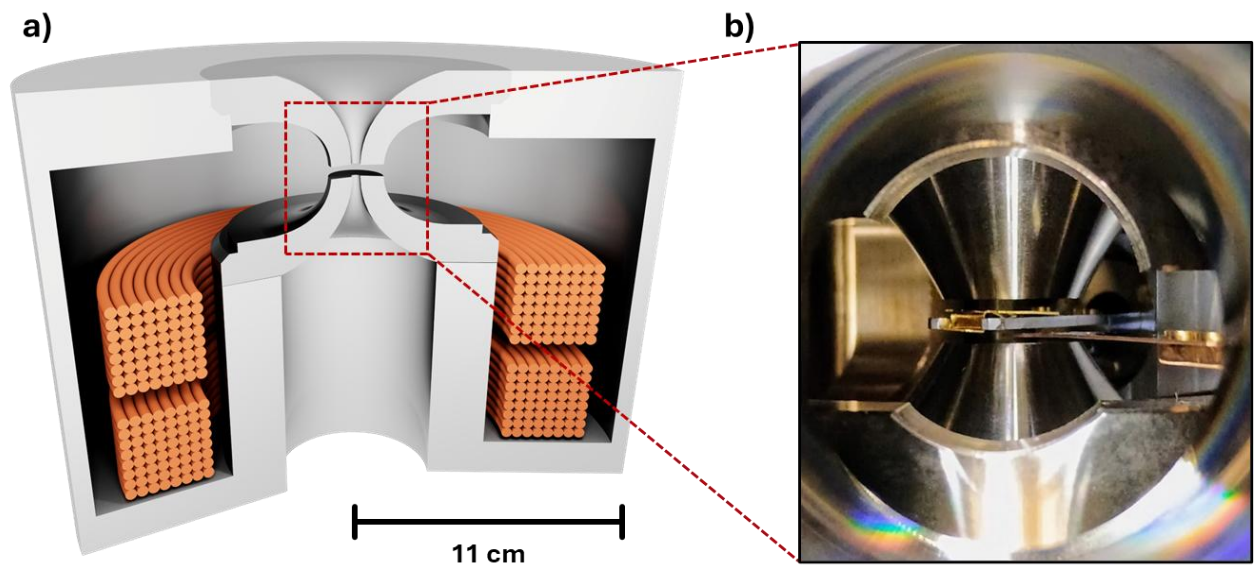

*Figure 1 (a) Illustration of a transmission electron microscope objective-lens; showing the copper wire coils, outer magnetic circuit, and pole-piece with pole-gap. (b) Photograph of the Dublin lens installed in a JEOL 1400/HR at the MRC-LMB. The specimen rod is visible extending from the rear on the right. The objective aperture strip extends from the front on the right.*

Such TEM objective lenses are normally designed to minimise Cs [8],[9]. Minimisation of Cs tends concurrently to minimise Cc as the two are coupled to the geometry of the pole piece [10],[11],[12],[13]. Notable exceptions include lenses designed for aberration-corrected systems [14],[15], which are normally operated with room-temperature specimens. Cc values for previous lenses range from 0.6 mm to 4.0 mm [16]. Designs with Cc less than 2 mm are rare and inevitably are designed for room temperature or ultra-high-vacuum columns [14]. Microscopes designed for cryoEM have Cc in the range 2-3 mm (e.g. Krios, CRYO-ARM, Polara) often to accommodate the wide gaps needed for a tight cryobox around the specimen while retaining the ability to tilt the grids as much as possible. Here, by eliminating the requirement for tilting the specimen, optimising specifically for Cc only and maintaining as wide a gap as possible, we have a new set of design constraints that are specific for improving single particle cryoEM.

**Methods**

*Approach*

Here we sought to design a lens suitable for accommodating a liquid nitrogen cooled anti-contaminator system and standard cryo-specimen holders without touching the pole-piece which is kept at room temperature.

*Design*

To evaluate the performance of many lens geometries in parallel, a genetic algorithm (GA) tailored specifically for bespoke pole-piece design was written [17]. The GA is a metaheuristic that relies on the theory of natural selection, wherein the members of a population with the most favourable traits are the most likely to survive and reproduce. The algorithm emulates this by preserving the best-performing geometries, creating mutants and crossovers of middling geometries, and discarding the worst candidates [18]. Performance was evaluated with a fitness function, specified by the user in accordance with the desired lens properties. Such a GA approach has previously been successfully implemented in magnetic and electrostatic lens design for objectives, electron guns and even MRI machines [19],[20],[21],[22],[23],[24].

Beginning with a Python-defined vectorised description of a stylised pole-piece geometry, various randomised mutations of the geometry were performed to create a population of candidates. Each candidate was used as the input for a magnetic flux simulation using COMSOL Multiphysics®. The resulting magnetic flux profile along the optic axis was used as the input to a Julia script to determine values of the chromatic aberration coefficient ($C_c$) as well as spherical aberration ($C_s$) and focal length. These values were substituted into a fitness function, in our case prioritising $C_c$, and the candidates were then ranked from best to worst. This ranking was used to create a new generation of geometries, say by eliminating the worst ranked half, followed by crossover and mutation of the remainder. The calculations continued in this iterative manner until the specified number of generations was reached.

Simulated aberration coefficients were calculated at a particular value of coil excitation just below magnetic saturation. This was obtained by evaluating the paraxial magnetic flux profile at two boundary excitation values to find the position of the image plane, which was above and below zero for the respective boundary values. Using these values as a starting point, paraxial flux analysis was performed iteratively at halfway points between boundary image plane positions until the coil excitation, for which the image plane lies at zero, i.e. the saturation excitation, was obtained. This value was used to calculate the final aberration coefficients utilised in the GA's fitness function.

For this work, the GA was driven by a fitness function which prioritised the minimisation of $C_c$ over $C_s$ or focal length. The fabrication-ready design resulted from the evaluation of 250 generations, each comprised of 250 candidates, with minor adjustments made to the curvature of the pole-pieces for machining feasibility. The simulated physical and aberration properties of the lens are displayed in Table 1, alongside the values measured and described later in this paper.

*Table 1. Design parameters and simulated properties (left) and the measured values (right).*

| Property (mm) | Specification / Simulation | Measured |
|---|---|---|
| Bore | 2.00 mm | 1.993 mm ($\pm$ 0.003) |
| Pole gap | 4.00 mm | 4.017 mm ($\pm$ 0.003) |
| $C_c$ | 1.015 mm | 0.97 mm ($\pm$ 0.02) |
| $C_s$ | 0.76 mm | 0.62 mm ($\pm$ 0.05) |
| Focal length | 2.43 mm | - |

*Fabrication*

Pole-pieces were fabricated in Ireland with assistance from Irish Manufacturing Research, a national centre of excellence and training in advanced manufacturing. Upper and lower poles were fabricated from an iron-cobalt alloy and a non-magnetic spacer placed between each pole. The complete pole-piece was heat treated to obtain optimal magnetic properties before dimensional analysis was performed using a coordinate measuring machine (CMM).

Before installation, the pole-piece was cleaned using an ultrasonic bath with sequential rinses of CMOS-grade acetone (x1) and isopropyl alcohol (x3) followed by air drying in a laminar flow clean hood. The original JEOL pole-piece was removed through the TEM's pole-piece access flange, and the new design was lowered in its place using a bespoke installation tool. The original goniometer and aperture strip were replaced and the column pumped.

*Testing*

The lens was installed on a standard JEOL 1400/HR microscope, which was equipped with two important modifications, namely a prototype hybrid pixel detector from DECTRIS and a Schottky FEG electron source from York Probe Sources.

The hybrid pixel detector was an experimental prototype that uses a gallium arsenide (GaAs) wafer as its sensitive layer with sub-50 μm pixels bump-bonded to the ASIC event counting architecture below, which consists of juxtaposed ASICs. The ion-pairs created by the electron track are collected by a voltage bias and converted into electron events that are read out at kHz frame rate. This gives the detector superb performance for its megapixel field-of-view, but a full description of its performance and properties are beyond the scope of this paper.

The field emission gun (FEG) from YPS is also a prototype. It represents a further improvement in terms of stability and reliability of the earlier design [25], used previously to demonstrate the promise of structure determination by cryoEM at 100 keV [4]. The most recent improvement involves the use of an improved high voltage ceramic material and improved mechanical alignment designs in the FEG.

After installation of the lens, the illumination system was realigned by adjustment of the beam tilt to minimise the image motion by observing the rotation centre using the objective lens current wobbler. A Zemlin tableau would be useful to provide further characterisation and is planned soon. [26].

Cc measurement was made by recording a series of images at a nominal magnification of 300 kX corresponding to a pixel size of 0.74 Å at the specimen, which was calibrated from the (111) fringes of the gold nanoparticles at 2.35 Å. The accelerating voltage was changed to be 30 V above and 30 V below the 100,000 V operating value in 30 V steps in both directions to eliminate any possibility of hysteresis. As in the earlier work [4], the voltage calibration was checked. The slope of defocus versus accelerating voltage gives Cc directly (Fig. 2a), using the formula Cc = (ΔF/ΔE)*E, where (ΔF/ΔE) is the slope and E is the accelerating voltage, and was found to be 0.97 mm.

Cs was estimated from a 300 kX magnification image (pixel size of 0.74 Å) of gold on carbon using the same C/Au/C sandwich specimen as used for Fig. 3b, but with an exposure time of 40 s, corresponding to 256,000 frames. Multiple sub-frame (2048) summations were then aligned by cross-correlation to compensate for slight specimen drift during the full exposure to give a final image that showed good resolution in all directions. Small amounts of differential magnification or astigmatism were then eliminated computationally to give the radial power spectrum shown in Fig. 2b. The values of defocus, Cs and amplitude contrast were then minimised computationally to produce the best agreement between the observed and fitted radii of the contrast transfer function (CTF) zeroes.

The two images shown in Fig. 3 were recorded with slightly higher magnifications (nominally 400 kX, with a calibrated pixel size of 0.55 Å at the specimen). Fig. 3a was made by adding together two images of amorphous platinum/iridium (Agar Scientific) recorded 40 seconds apart during which the specimen had drifted by about 35 Å, creating an electron imaging equivalent of the Young's fringe experiment [27]. The fringes show good resolution beyond 2 Å. Fig. 3b shows the FFT of an image of a specimen of gold nanoparticles sandwiched between thin films of amorphous carbon. This was made by floating carbon off mica, then evaporating gold followed by a final coating of carbon to create a C/Au/C sandwich in which the gold domain orientations were more stable than on a single layer. The real-space image is shown in Supplemental Figure S1.

## Results

We determined Cc at 100 keV by measuring the slope of defocus versus accelerating voltage (Fig. 2a). Cc was 1.01 mm with the stage at a height of Z=350 μm (Obj current 3.35 A). Cc was 0.97 mm (± 0.02) with the stage at the lower height of Z=150 μm (Obj current 4.02 A).

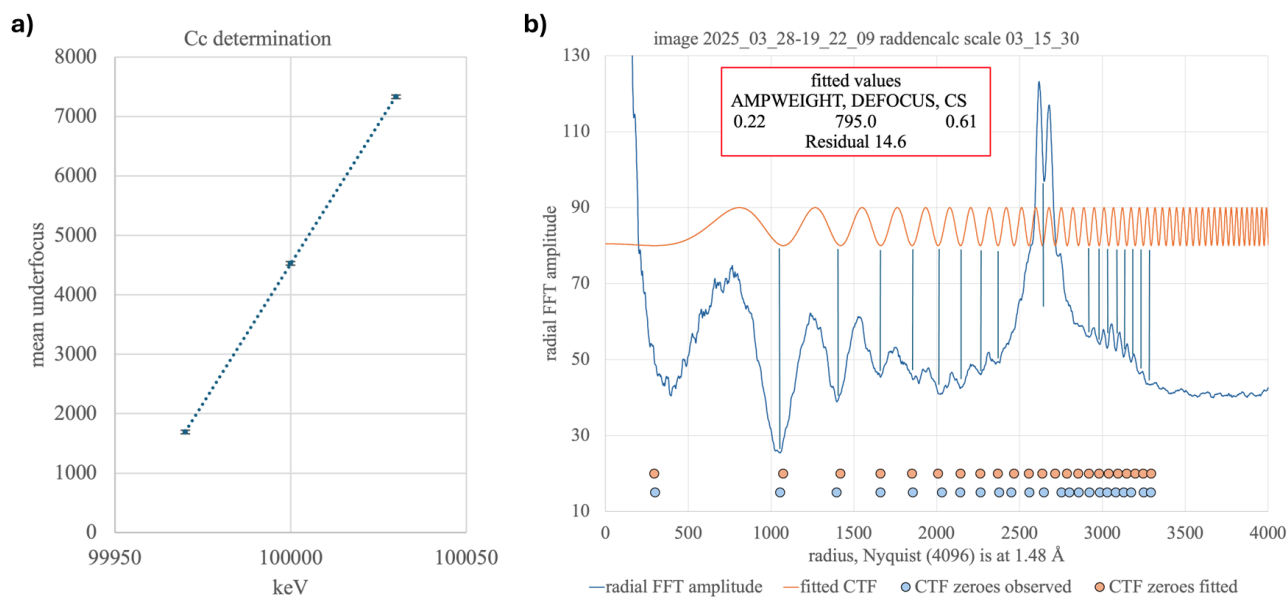

*Figure 2* (a) Measurement of Cc from change in underfocus as a function of electron energy; error bars on defocus are small. (b) Measurement of Cs by least-squares fitting of three CTF parameters shown in the box. The specimen was a carbon/gold/carbon sandwich.

The JEOL 1400 lens power supply did not allow enough current to obtain an in-focus image with the stage height lower than Z=150 μm (Objective lens current 4.10 A is the current maximum), but this could be easily fixed with simple changes to the lens power supply.

Cs was estimated by fitting the CTF to the Fourier transform of an image of sputtered gold sandwiched between two thin films of amorphous carbon. Cs was found to be 0.61 mm, as shown in Fig. 2b. The agreement between the observed and calculated CTF zeroes was not perfect but there were two mitigating factors, one experimental and one theoretical, that could explain the deviation. Experimentally, the C/Au/C specimen was made by floating the first layer of carbon off mica so it should be very flat. Gold nanoparticles were then sputtered onto the carbon film before coating with a second layer of carbon. Thus, the gold may have a slightly different height (possibly by ~50 Å) than the carbon. A defocus change from 823 Å to 800 Å would cause a shift of the CTF zeroes by the small amount observed. Theoretically, it is also known that the amplitude contrast for elastic scattering from gold is much higher than for carbon. For example, Erickson and Klug showed [28] that uranyl acetate had 35% amplitude contrast, whereas carbon has only about 10%. Such a variation would also cause a shift in the CTF zero positions for the gold fringes. A similar value for Cs of 0.62 (± 0.05) mm was obtained using images of a pure carbon film, with closer fit but over a smaller range because the Thon rings from the carbon film extended only to about 2.5 Å.

The performance of the lens was demonstrated by recording images of amorphous PtIr displayed using the Young's fringe approach (Fig. 3a, see legend) and gold sputtered on a thin film of amorphous carbon (Fig. 3b), showing the lattice spacings expected from the gold nanoparticles.

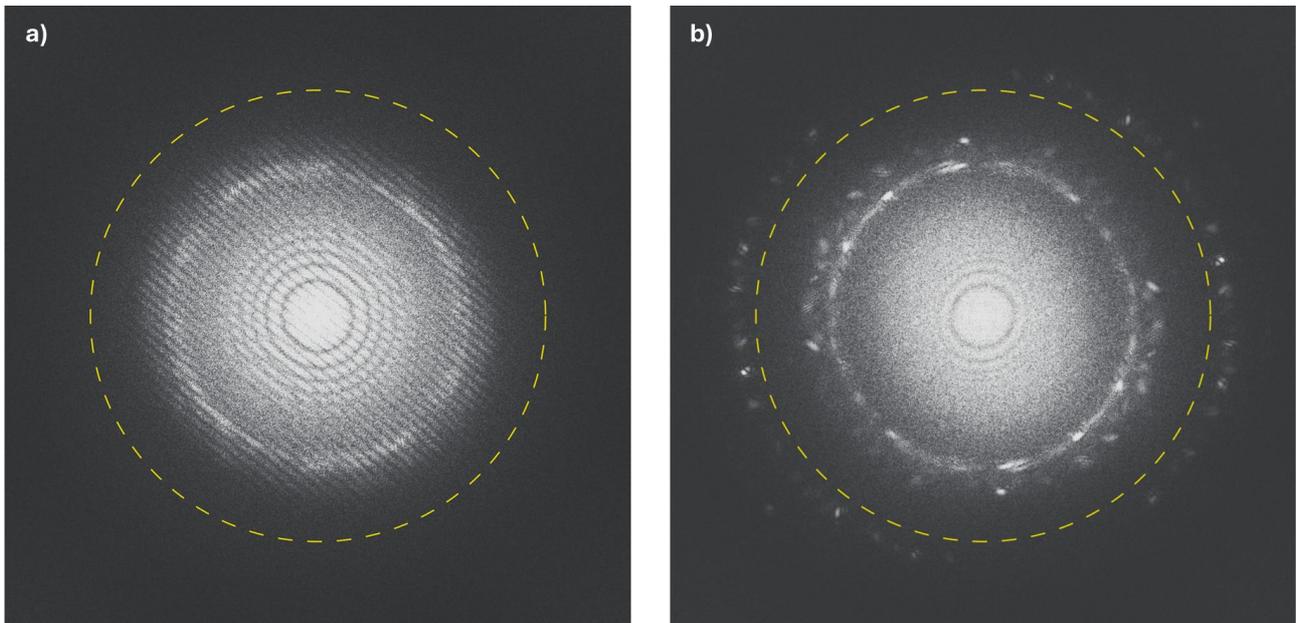

*Figure 3* (a) Young's fringes FFT of the sum of a pair of PtIr images extending well beyond 2 Å. The images added together were 2025_03_06-21_31_03 and 2025_03_06-21_31_42. (b) FFT of gold in a carbon sandwich showing the 2.3, 2.0, 1.4 and 1.2 Å lattice fringes from Au (111), (200), (220), and (311) spacings in image 2025_03_19-17_55_02. Images were recorded at a nominal magnification of 400 kX using a DECTRIS prototype detector, with a calibrated pixel size of 0.55 Å at the specimen. The resolution at the edge of the image is 1.1 Å and the dashed circle is at 1.5 Å.

**Discussion**

Most previous objective lens designs gave priority to reducing Cs. Here we prioritised reducing Cc, since in cryoEM Cs is already fully included in CTF correction for the defocused bright-field images that are essential to produce enough contrast. CryoEM also needs a wide enough pole-gap to allow space for an anti-contaminator. The lens described here with a 2 mm bore and 4 mm gap fulfils these goals.

Note also that the narrow gap means that specimen tilts beyond a few degrees would not be possible, but single-particle cryoEM does not need specimen tilt. Alongside these requirements, it remains important to have a stable cold stage and rapid cryo-transfer, which are features that are beyond the scope of this paper.

**Conclusions**

We have shown that it is possible to design, fabricate and test an objective lens for 100 keV microscopy that has a wide pole-piece gap, a reasonable bore, performance that agrees with expectations from simulations, and a resulting improvement of the signal at 2 Å resolution. We hope that this approach to designing a lens for cryoEM where minimum Cc is the key parameter while keeping costs down, will be adopted for all single-particle electron cryomicroscopes going forward.

**Acknowledgements**

GMA and LJ acknowledge Research Ireland grant 12/RC/2278 P2 (AMBER2), TA acknowledges Research Ireland grant EPSPG/2024/1994, and LJ acknowledges Research Ireland grant URF/RI/191637. The authors would like to thank Maixent Cassagne for help testing the GA code during his internship, and Dr. Tomáš Radlička of the Czech Academy of Sciences for developing the Julia package with which the simulated aberration coefficients were calculated. The Trinity authors would like to thank Irish Manufacturing Research for their assistance in developing the high-precision manufacturing and dimensional metrology. CJR, GM and RH thank the Wellcome Trust for award of grant No. 220526/B/20/Z, together with valuable financial support from Astex and AstraZeneca. This work has also been supported by the Medical Research Council under Grant No. MC_UP_120117 (C.J.R.). We acknowledge early support for development of the FEG from Innovate UK Grant No. 103806. We would like to thank Karl Gaff for the illustration of the objective lens, P. Nellist and N. Unwin for helpful discussions.

**Declaration of Interests**

The authors declare the following interests: LJ is the CEO and co-founder of turboTEM Ltd. who, amongst other activities, manufacture bespoke TEM pole-pieces. GMA is a former Trinity College researcher now employed by turboTEM Ltd. TA is a PhD student jointly funded by Research Ireland and turboTEM Ltd.

**Supplementary Information**

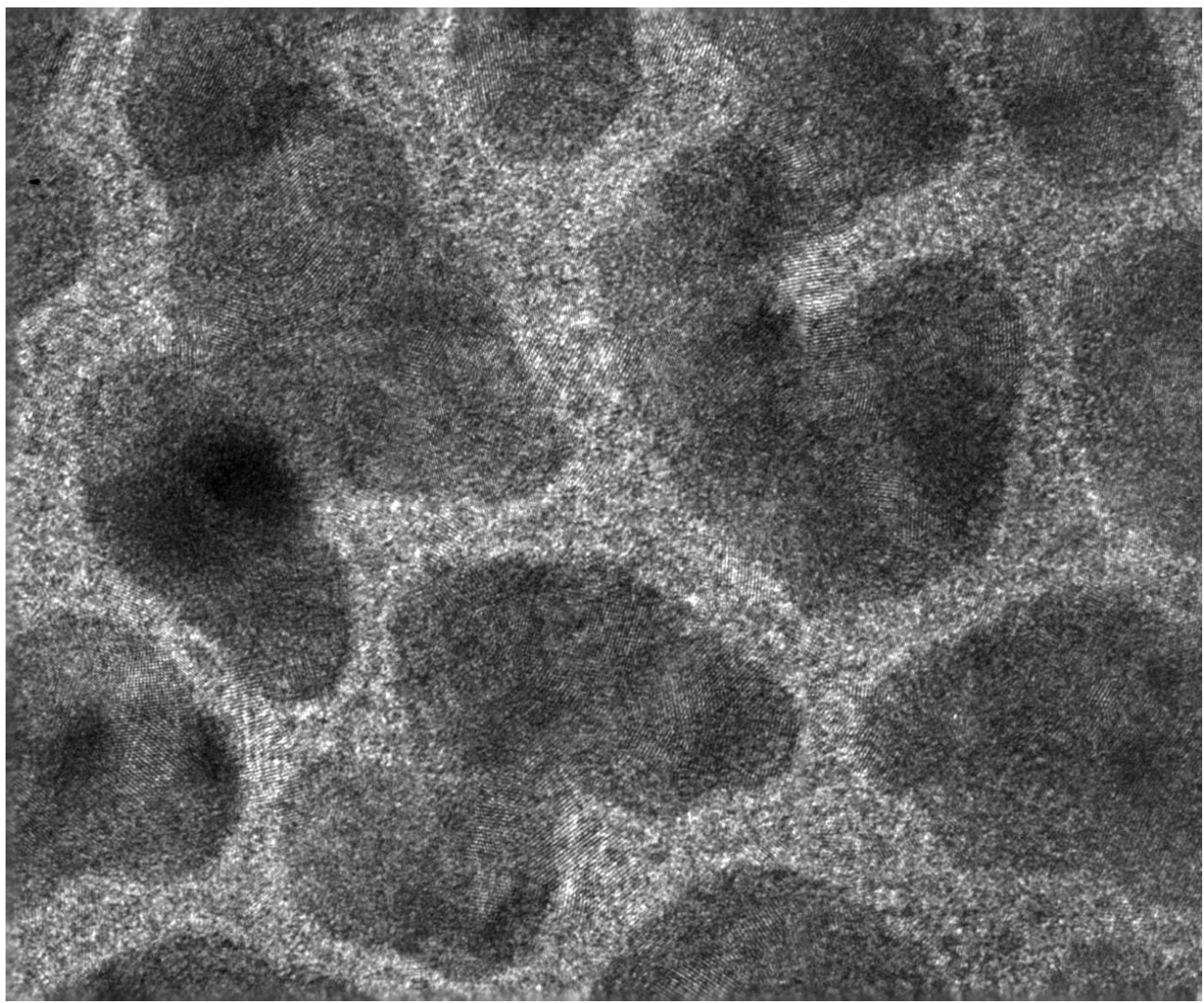

***Figure S1.*** *Image of gold in a carbon sandwich (file: 2025_03_19-17_55_02) showing clear lattice fringes. Images were recorded at a nominal magnification of 400 kX using a DECTRIS prototype detector, with a calibrated pixel size of 0.55 Å.*